\documentclass[%
 reprint,
 twocolumn,
superscriptaddress,
 amsmath,amssymb,
 aps,
 enumitem,
]{revtex4}

\usepackage{graphicx}
\usepackage{dcolumn}
\usepackage{bm}
\usepackage{textgreek}
\usepackage{gensymb} 
\usepackage[dvipsnames]{xcolor} 
\usepackage[normalem]{ulem} 
\usepackage{hyperref}

\usepackage{enumitem,amssymb}
\newlist{todolist}{itemize}{2}
\setlist[todolist]{label=$\square$}
\usepackage{pifont}
\begin{document}

\preprint{APS/123-QED}

\title{Droplets sit and slide anisotropically on soft, stretched substrates}

\author{Katrina Smith-Mannschott}
\affiliation{Department of Materials, ETH Z\"{u}rich, 8093 Z\"{u}rich, Switzerland.}%
 
 \author{Qin Xu}
\affiliation{Department of Materials, ETH Z\"{u}rich, 8093 Z\"{u}rich, Switzerland.}
\affiliation{
Department of Physics, The Hong Kong University of Science and Technology, Hong Kong, China}

\author{Stefanie Heyden}%
\affiliation{Department of Materials, ETH Z\"{u}rich, 8093 Z\"{u}rich, Switzerland.}%

\author{Nicolas Bain}
\affiliation{Department of Materials, ETH Z\"{u}rich, 8093 Z\"{u}rich, Switzerland.}

\author{Jacco H. Snoeijer}
\affiliation{Physics of Fluids Group, Faculty of Science and Technology,
Mesa+ Institute, University of Twente, 7500 AE Enschede, The Netherlands
}

\author{Eric R. Dufresne}
\affiliation{Department of Materials, ETH Z\"{u}rich, 8093 Z\"{u}rich, Switzerland.}

\author{Robert W. Style}
\affiliation{Department of Materials, ETH Z\"{u}rich, 8093 Z\"{u}rich, Switzerland.}%
 \email{robert.style@mat.ethz.ch}

\date{\today}

\begin{abstract}
Anisotropically wetting substrates enable useful control of droplet behavior across a range of   applications.
Usually, these involve chemically or physically patterning the substrate surface, or applying gradients in properties like temperature or electrical field.
Here, we show that a flat, stretched, uniform soft substrate also exhibits asymmetric wetting, both in terms of how droplets slide and in their static shape.
Droplet dynamics are strongly affected by stretch: glycerol droplets on silicone substrates with a 23\% stretch slide 67\% faster in the direction parallel to the applied stretch than in the perpendicular direction.
Contrary to classical wetting theory, static droplets in equilibrium appear elongated, oriented parallel to the stretch direction.
Both effects arise from droplet-induced deformations of the substrate near the contact line.
\end{abstract}

\maketitle


Liquid droplets  isotropically wet most solid surfaces.
However,  living organisms have evolved a large variety of anisotropic surfaces that provide novel functionality, \cite{park01,zheng2010directional,zheng07}.
These have inspired artificial anisotropic surfaces, which show promise for diverse applications from microfluidics  to fog harvesting  \cite{xia12, squi05, gleiche01,guersoy17}. 
Anisotropically-wetting surfaces can be created by different approaches, including chemically patterning surfaces with surface-energy gradients to drive droplets from high to low energy regions \cite{chau92} or with stripes and other patterns to cause asymmetric spreading and pinning of droplets \cite{morita2005macroscopic}.
Similarly, micro-patterning surfaces with bumps, pillars, or various other features also drive direction-dependent wetting \cite{chu2010uni,malvadkar2010engineered,dressaire2009thin,kusumaatmaja2008anisotropic,dressaire2009thin}.
All these surfaces involve significant efforts in fabrication.

Here, we show that droplets anisotropically wet slabs of soft material that have been anisotropically stretched.
Furthermore, droplets respond anisotropically to applied forces,  sliding faster in one direction than another. 
These effects contradict classical wetting theory.
We show that they are governed by an anisotropic response of the wetting ridge, the microscopic deformation of the substrate localized to the contact line (\emph{e.g.} \cite{peri08,styl13,bard18}).

\begin{figure}[b]
    \centering
        \includegraphics[width=.7 \columnwidth]{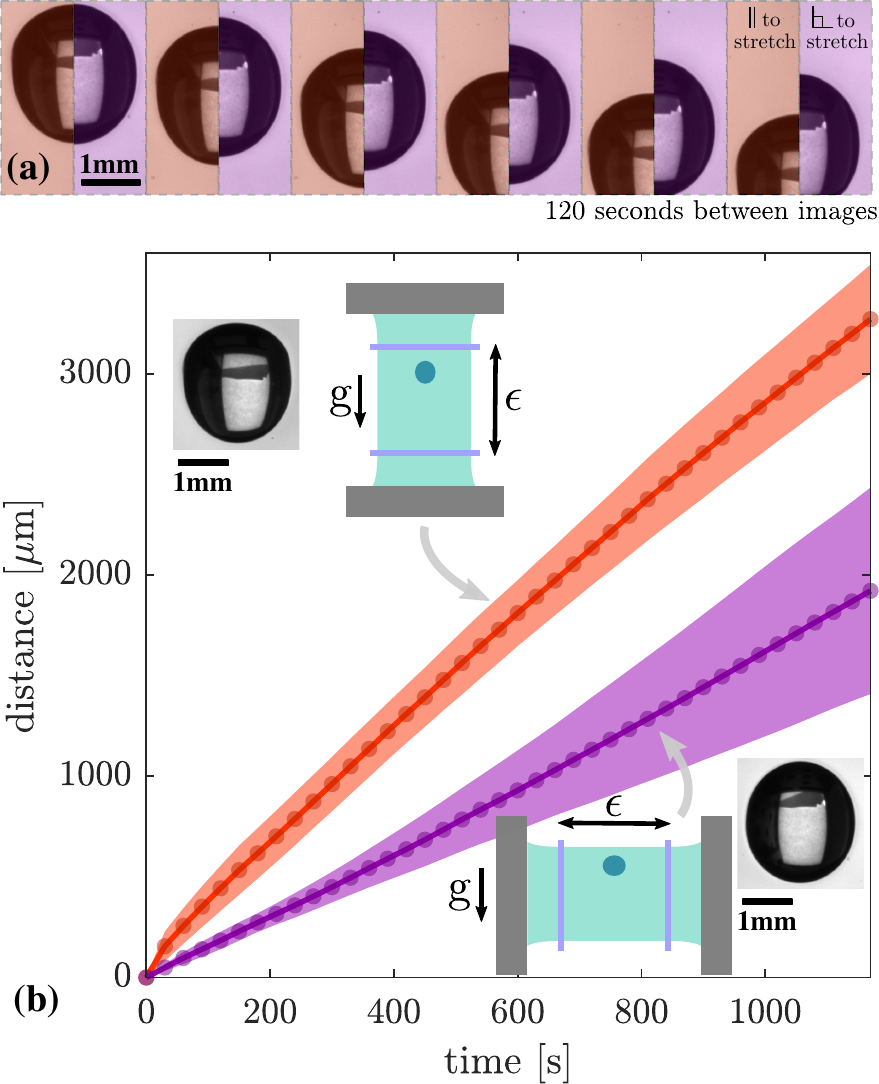}
    \caption{Droplet dynamics slide anisotropically on stretched substrates. 
    a) Droplets sliding parallel to the stretch (red) direction move faster than those sliding in the perpendicular (purple) direction and have clearly different shapes.
    b) Measured droplet positions vs time for sliding in the two different directions.
    This shows the average for 9 droplets sliding parallel to stretch and 7 droplets sliding perpendicular to stretch. $\epsilon=23\%$. 
    Shaded areas indicate one standard deviation to either side of the average.
    Note that the speed is essentially constant during our observations. Droplets do not undergo a transition to fast, lubricated sliding, indicating that un-crosslinked silicone chains in the substrate do not play a role here \cite{hour17}.
    }
    \label{fig:slidingdrops}
\end{figure}

We study the effect of strain on wetting behavior by investigating the behavior of glycerol droplets on soft, 1.0~mm thick, silicone gels (CY52-276, Dow Corning) with a Young's modulus of $E=6$~kPa, supported on a much stiffer, stretchable membrane \cite{xu17}.
In equilibrium, on both the stretched and unstretched substrates, these droplets have a  contact angle of 91$^\circ$ \cite{xu17}.
We cut narrow strips from the substrate and uniaxially stretch these on a home-built stretching device. 
To ensure complete flatness, which we verify with interferometry in the Supplement, we placed sections of 1mm-diameter glass capillaries on the surface of the gel to suppress residual, longitudinal wrinkles (\emph{c.f.} Fig. \ref{fig:slidingdrops}).
Strain, $\epsilon$, was calculated either from imaging markings on the samples or from the displacement of the stretching device (see Supplement for details).

\begin{figure*}
    \centering
    \includegraphics[width=0.80 \textwidth]{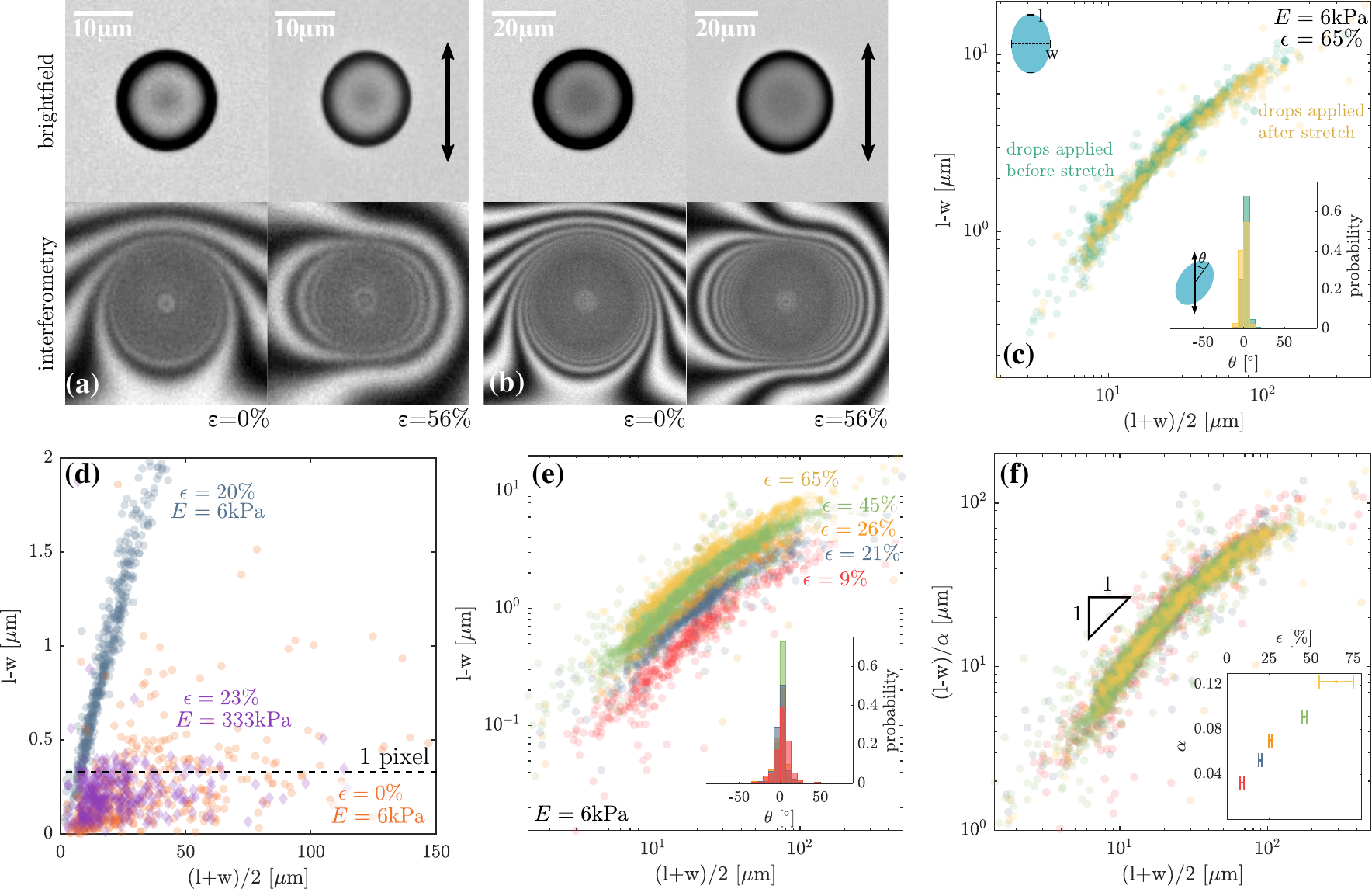}
    \caption{
 Droplet shapes become elliptical on soft, stretched substrates.
     a,b) Brightfield (top) and interferometry (bottom) images for two droplets of different sizes, when the substrate is relaxed (left) and after 50\% stretch (right). The stretch direction is indicated by the arrows.
    c) There is no observable contact line hysteresis for droplets. Droplets are the same shape regardless of whether they are applied to a soft, stretched substrate before it is stretched (green) or after (yellow).
    The inset histograms show the alignment of droplets with the stretch direction.
    d) Droplets are essentially round (to pixel resolution) on unstretched, soft silicone (red circles) and stretched, stiff silicone (purple diamonds). The blue data shows droplet shapes on soft silicone with a similar stretch for comparison.
    e) $l-w$ for glycerol  droplets on soft silicone substrates increases with droplet size and stretch. The inset histograms show the alignment of droplets with the stretch direction.
    f) The data from e) collapses onto a single curve, when each data set is divided by the average slope, $\alpha$, of the data for $(l+w)/2<20$\textmu m.
    The inset shows this collapse parameter $\alpha$ as a function of $\epsilon$. }
    \label{fig:megapic}
\end{figure*}

Recent experiments found that contact line motion depends on stretch for similar  surfaces \cite{snoeijer2018paradox}, but did not explore potential anisotropy in the response.  We find that droplets slide faster in the direction of applied stretch.
As shown in Fig. \ref{fig:slidingdrops}a,
 5\textmu L glycerol droplets slide down a vertical substrate with a uniaxial stretch of 23\%, oriented either parallel  (red) or perpendicular (purple) to gravity.
This experiment was for 7-9 droplets in each orientation and the average trajectories are plotted  in Fig. \ref{fig:slidingdrops}b.
Droplets sliding parallel to the stretch direction are 67\% faster than those sliding perpendicular to the stretch direction.
Interestingly, they also have clearly different shapes, with the faster droplets appearing to have a marked asymmetry between trailing and leading edges.
Thus, moving millimetric droplets are sensitive to underlying stretch of soft materials.

The static shapes of droplets are also sensitive to substrate deformation. 
Droplets tend to elongate in the direction of applied stretch, as shown in 
in Fig. \ref{fig:megapic}a,b.
There, images of two differently-sized droplets before and after $\epsilon=50\%$ stretch are imaged with  brightfield and interference microscopy.
In the unstretched case, droplets all remain completely circular, while in the stretched case, there is a small elongation along the stretch direction.
While the effect is subtle in brightfield, there is a stark difference when the same droplets are imaged with an interferometric objective (bottom row), which shows changes to the substrate profile around the droplets.
In the unstretched case, the interferometry fringes around the droplets are essentially symmetric and angle-independent (the top/bottom asymmetry results from a slight tilt of the substrate). 
However, in the stretched case, we see a strongly asymmetric, lobed pattern that extends perpendicular to the stretch direction.

Stretch-induced changes of shape could be a result of contact-line pinning and have been observed on surfaces with strong ($>15^\circ$) contact-angle hysteresis \cite{good1971anisotropic}.
However, previous observations have found that this silicone has negligible static contact-angle hysteresis  \cite{styl13b,snoeijer2018paradox} and shows no evidence of a non-zero roll-off angle   \cite{karp16}.
To minimize possible hysteresis,  we further equilibrate droplets for at least one hour before any measurements.
We confirm the effectiveness of this approach by comparing the final shape of droplets that are applied to the substrate before or after it is stretched.
We quantify droplet shape by plotting the difference between the long and short axis lengths of over 1400 droplets, $l-w$, versus their average  radius, $(l+w)/2$  in Fig. \ref{fig:megapic}c.
The shapes of droplets deposited before stretch (green) are identical to those deposited after stretch (yellow).
This confirms that that there is essentially no hysteresis at these timescales on this substrate.
In both cases, the long axis of the droplet is always aligned with the 
the  direction of stretch, as shown in the inset of Fig. \ref{fig:megapic}c.

This droplet-stretching behavior is limited to substrates with very low elastic moduli.
We compare the shapes of droplets on the soft silicone  with droplets on a 1.1mm-thick slab of much stiffer silicone  ($E=333$kPa, made following \cite{kim20}), as shown in Fig. \ref{fig:megapic}d.
While over 600 droplets on the soft substrate at $\epsilon=20\%$ (navy blue) show a pronounced increase of anisotropy with size, 345 droplets on the stiff substrate at a similar strain of $\epsilon=23\%$ (purple) show no systematic variation in shape, with most of the  recorded values of $l-w$ falling below the  resolution of the imaging system.
Indeed, the shapes of droplets on stretched stiff substrates are indistinguishable from those on unstretched, soft ones, shown in red.
This is consistent with previous reports, which found no observable change to droplet shape when placed on stiffer elastomers with different stretches \cite{schu18}.

These results suggest that the  equilibrium shape of droplets on soft substrates depends on their size and the applied strain.
To that end, we quantified the shapes of over 2150 droplets from 2 to 40 \textmu m in radius over a range of uniaxial strains from 9-65\%, as shown in Fig. \ref{fig:megapic}e.
As seen previously, $l-w$ always increases monotonically with droplet size with the long axis of the droplet aligned with the stretch direction (see inset).
For smaller droplets, the asymmetry, $l-w$, increases linearly with the droplet size.
Equivalently, small droplets at the same applied stretch all have the same  aspect ratio, $l/w$.
At larger strains, the asymmetry increases more slowly with droplet size and the droplets become more circular  (\emph{i.e.} $l/w \rightarrow 1$).
This trend is identical for all non-zero strains, as highlighted in Fig. \ref{fig:megapic}f.
There each data set is normalized by the slope, $\alpha$ of the small-droplet data (radius $< 20$\textmu m).
Not only does this collapse the data onto a single curve for the small droplets, as it must, but it also collapses the transition to sublinear behavior for large droplets.
This suggests a  shift in the underlying physics occurring at a characteristic length scale around 30 \textmu m.
Note that the scale factor, $\alpha$, is not simply proportional to the applied strain,  $\epsilon$ (see the inset).
Thus the droplet response is not linear, and scaling the asymmetry by the applied strain, as in \cite{styl15}, does not collapse the data.

What drives the anisotropy? The classical theory of wetting for flat, uniform surfaces (\emph{i.e.} Young-Dupr\'{e}) only depends on scalar surface energies.
Even if the surface energy were strain-dependent, it would still predict circular contact lines.
A clue to the missing physics is given in Fig. \ref{fig:megapic}f, where the transition away from a constant aspect-ratio regime always happens at the same length scale, independent of $\epsilon$.
This is the hallmark of an elastocapillary effect \cite{styl17}, which are typically characterized by a shift in behavior around one of the elastocapillary lengths, $\Upsilon/E$ or $\gamma_{lv}/E$, where $\Upsilon$ is the solid's surface stress and $\gamma_{lv}$ is the droplet's surface tension (\emph{e.g.} \cite{lester1961contact,zimb07,jago12,mora14,chak13,styl17}).
For our soft substrates, these length scales are $\mathcal{O}$(10~\textmu m) \cite{styl13}.
Below this length scale, we generically expect surface properties to dominate behavior over bulk elasticity.

In the limit of small droplets, we therefore expect droplet shape to be determined by a balance between the surface stresses of the substrate, and the surface tension of the droplet, in a manner independent of the droplet size.
If the surface stresses are strain independent and thus isotropic, we might expect the droplet to retain its circular shape.
However, if surface stresses are strain-dependent, as reported in \cite{xu17,xu18}, uniaxial stretch would induce anisotropic surface stresses and likely result in non-spherical droplet shapes, like the ones seen in the small-droplet limit of our experiments.
In the limit of large droplets, much bigger than the elastocapillary length scale, we expect to recover the classical wetting behavior of Young-Dupr\'{e}, and so droplet asymmetry should vanish.
This is consistent with the lack of asymmetry of droplets on the stiffer substrate in Fig. \ref{fig:megapic}d, where the elastocapillary lengths are of order $\mathcal{O}(100~\mathrm{nm})$.

\begin{figure}[t]
    \centering
    \includegraphics[width=1 \columnwidth]{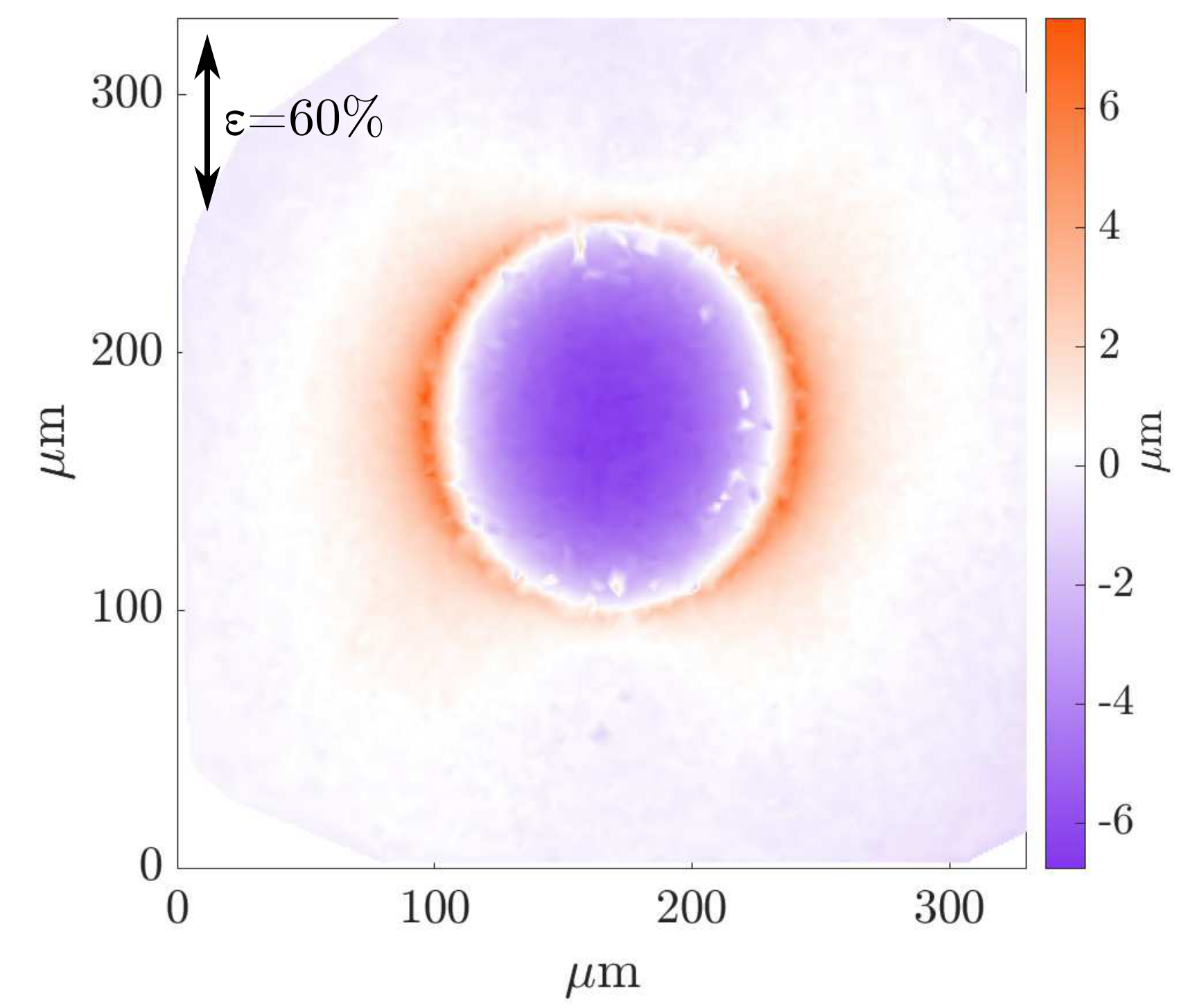}
    \caption{The surface profile under a droplet on a stretched substrate ($\epsilon= 60\%$) is very asymmetric. 
     We measure the height profile of fluorescent nanobeads on the surface of the silicone gel under a glycerol droplet with confocal microscopy (e.g. \cite{styl13}).
    The color is an interpolated surface profile.}
    \label{fig:confocal}
\end{figure}

Elastocapillary effects revolve around substrate deformations.
Thus, if droplet elongation is  elastocapillary in origin, we expect  anisotropic substrate deformations around the droplet.
This is hinted at by the interferometry images in Fig. \ref{fig:megapic}a, but is properly visualized via confocal measurements of the surface profile under the droplet, as in Fig. \ref{fig:confocal}.
This clearly shows the wetting ridge at the contact line and how it is significantly shorter in the direction of the applied $\epsilon=60\%$ uniaxial stretch.
A similar anisotropy of the wetting ridge  was previously  reported \cite{xu18}, and attributed to the presence of a strain-dependent surface stress of the substrate.

This ridge anisotropy likely underlies the asymmetric static droplet shapes. 
Previous work has shown that the presence of a wetting ridge reduces the energy of a droplet, much like a negative line tension, predominantly by covering up droplet surface area \cite{lubb14}.
Taller ridges lower the energy more, so a droplet will extend in the direction of stretch, as then a larger fraction of its contact line will feature a taller wetting ridge.
However, in this static case, making a predictive theory is challenging, due to the complex, non-symmetric geometry.

However, for the dynamic case, we can quantitatively link the asymmetric dynamics of large droplets to the microscopic ridge anisotropy.
Previous work has shown that the speed of droplets sliding on soft substrates, $v$, is set by viscoelastic dissipation at the wetting ridge \cite{carr96,karp15,kaji13,zhao18, long96}.
We build upon these results to explain our data.
Let $G'(\omega)$ and $G''(\omega)$ be the  frequency-dependent, substrate storage and loss moduli respectively.
We approximate the ridge in the sliding direction as a triangle with height and width
$h_r$,$w_r$ respectively.
so that the characteristic sliding  frequency, $\omega_c=v/w_r$ \cite{karp15} and the 
 characteristic strain in the ridge is $\epsilon_r\sim h_r/w_r$. Then, balancing the energy dissipation rate in the ridge with the gravitational energy released as the droplet slides down the substrate, we find:
\begin{equation}
     Mg v=G''(\omega_c)V_r \epsilon_r^2 \omega_c \sim G''\left(\frac{v}{w_r}\right) R h_r^2 \frac{v}{w_r},
     \label{eqn:diss_balance}
\end{equation}
where $Mg$ is the weight of the droplet and $V_r\sim w_r^2 R$ is the volume of the material deformed under the ridge.
Thus, we see that $v$ is intimately linked to the wetting ridge shape.

We can not directly evaluate this expression without tricky measurements of the shape of the dynamic wetting ridge, $h_r(v),w_r(v)$ (c.f. \cite{van20}), but these can be approximated at slow speeds by $h_r(0),w_r(0)$. 
For viscoelastic solids, there is a characteristic frequency, $\omega^*$, below which $G'(\omega)$ is effectively constant, and $G'(\omega) \gg G''(\omega)$.
To leading order, when $\omega_c <\omega^*$, the dissipation can thus be computed from the static ridge geometry.
Using $w_r\sim 10~\mu$m (from Fig. \ref{fig:confocal}), and $v\sim 1~\mu\mathrm{m/s}$ (from Fig. \ref{fig:slidingdrops}),
we estimate $\omega_c\sim 0.1~\mathrm{s}^{-1}$, 
 much smaller than $\omega^* \sim 1~\mathrm{s}^{-1}$ \cite{karp15}.
Thus, we can use previous static measurements  \cite{xu17,xu18} of the solid ridge-tip angle, $\theta_s(\epsilon)$, to evaluate Equation (\ref{eqn:diss_balance}).
We incorporate this via the scalings
$h_r\sim \gamma_{lv}/E$, and $w_r\sim h_r \tan (\theta_s/2)\sim \gamma_{lv}/(E \cos(\theta_s/2))$.
The last approximation, which comes from assuming small ridge slopes, is used to match previous analytic expressions that accurately predict the dynamic contact angles of moving contact lines on similar substrates \cite{snoeijer2018paradox,karp15}.

Finally, we use the fact that our silicone substrate has a common power-law rheology, $G''(\omega)\approx E(\omega \tau)^n/3$, where $\tau=0.13$~s and $n=0.51$ (e.g. \cite{karp15,snoeijer2018paradox,xu20}), to obtain 
\begin{equation}
    v=\frac{\gamma_{lv}}{E \tau} \left(\frac{Mg}{\lambda\gamma_{lv} R \cos^{n+1}(\theta_s/2)}\right)^{1/n},
    \label{eq:v}
\end{equation}
where $\lambda$ is a dimensionless, $O(1)$ prefactor. 
Thus we see that when droplets slide perpendicular to tall, pointy ridges with small $\theta_s$, they move slowly due to increased dissipation.
However, when the ridge is small and flat, and $\theta_s$ is large, dissipation is reduced, and droplets travel faster (c.f. Fig. \ref{fig:slidingdrops}). 

With no free parameters, this model accurately predicts the  relative increase of the sliding speed in the direction parallel to the applied stretch, as reported in Fig. \ref{fig:slidingdrops}. 
To evaluate Eq. \ref{eq:v}, 
we interpolate previous experimental observations of $\theta_s(\epsilon)$ \cite{xu18,heyden2020contact}, giving  $\theta_s=122.4^\circ$ and $131.8^\circ$ for ridges oriented perpendicular and parallel to the stretch direction, respectively.
This gives that the speed parallel to stretch, $v_{\parallel}$, should exceed the speed in the perpendicular direction, $v_{\perp}$ by 63\%, in good agreement  with the experimentally observed increase of 67\%.
Furthermore, we find that taking $\lambda=18.8\approx 6 \pi$ matches observed sliding speeds, where we use that $R=1$ mm, $Mg=20.6~\mu$N, and $\gamma_{lv}=41$ mN/m \cite{xu17}.

Note that the stretch-induced anisotropy of droplets seen here on thick, soft substrates should not be confused with a similar response for droplets on very thin, stretched, elastic membranes ($\mathcal{O}(100~\mathrm{nm)}$) \cite{schu15,schu17}.
There, the anisotropic droplet shape is understood to arise from a different mechanism: a balance of droplet surface tension with thickness-integrated elastic stresses  \cite{kumar2020stresses}.
Furthermore, in that case, anisotropy is only expected for droplets that are much larger than a characteristic length-scale determined by the membrane rigidity and the thickness-integrated stresses in the sheet, which must be  $\lesssim \gamma_{lv}$.

In conclusion, we have shown that simple, flat, soft, stretched substrates show asymmetric wetting, with a particularly strong effect on wetting dynamics.
This is significant, as most asymmetrically-wetting substrates are more complex.
Substrate stretch could represent a novel approach to droplet control, that can be easily tuned \emph{in situ}.
Thus, it could modulate processes including adhesion \cite{butl19}, condensation \cite{soku10}, and coalescence \cite{chudak2020escape}, and drive movement along strain gradients (\emph{c.f.} \cite{marq19}).
Our observations cannot be explained with a classical, macroscopic description of wetting.
Instead they are elastocapillary phenomena, driven by microscopic, asymmetric deformations of the substrate under droplets.
Despite recent progress in theory, there is still no consensus on what controls the size and shape of these deformations under macroscopic stresses.
For example, one school of thought proposes that the contact-line ridge shape only depends on a force balance between the surface stresses of the substrate and the droplet surface tension (e.g. \cite{xu17,xu18,pandey2020singular,heyden2020contact}),  while other, recent works have suggested a role for bulk elastic stresses (e.g. \cite{masu19, wu2018effect,liang2018surface}).
The theoretical description that we have presented is independent of these considerations.
However, we hope that an analysis of our static experimental results will help pinpoint the key physics.

\begin{acknowledgments}
This research was funded by the Swiss National Science Foundation (grant number 200021-172827).
\end{acknowledgments}

\end{document}